\documentclass[epsf,prb,twocolumn,showpacs,nofootinbib,nosuperscriptaddress]{revtex4-1}

\usepackage[pdftex]{graphicx}
\usepackage{dcolumn}
\usepackage{bm}
\usepackage{epsfig}
\usepackage{latexsym}
\usepackage{amsmath}
\usepackage{amsfonts}
\usepackage{amssymb}
\usepackage{float}
\usepackage{caption}
\usepackage{graphicx}
\graphicspath{{images/}}
\usepackage{subcaption}
\usepackage{color}
\usepackage{array}
\usepackage{framed}
\usepackage{soul}

\begin{document}

\title{Dynamical scar states in driven fracton systems}

\author{Shriya Pai}
\affiliation{Department of Physics and Center for Theory of Quantum Matter, University of Colorado, Boulder, CO 80309}

\author{Michael Pretko}
\affiliation{Department of Physics and Center for Theory of Quantum Matter, University of Colorado, Boulder, CO 80309}

\date{\today}

\begin{abstract}
One-dimensional fracton systems can exhibit perfect localization, failing to reach thermal equilibrium under arbitrary local unitary time evolution.  We investigate how this nonergodic behavior manifests in the dynamics of a driven fracton system, specifically a one-dimensional Floquet quantum circuit model featuring conservation of a $U(1)$ charge and its dipole moment.  For a typical basis of initial conditions, a majority of states heat up to a thermal state at near-infinite temperature.  In contrast, a small number of states flow to a localized steady state under the Floquet time evolution.  We refer to these athermal steady states as ``dynamical scars," in analogy with the scar states observed in the spectra of certain many-body Hamiltonians.  Despite their small number, these dynamical scars are experimentally relevant due to their high overlap with easily-prepared product states.  Each scar state displays a single agglomerated fracton peak, in agreement with the steady-state configurations of fractonic random circuits.  The details of these scars are insensitive to the precise form of the Floquet operator, which is constructed from random unitary matrices.  Rather, dynamical scar states arise directly from fracton conservation laws, providing a concrete mechanism for the appearance of scars in systems with constrained quantum dynamics.
\end{abstract}
\maketitle

\normalsize

\emph{Introduction}.    Quantum many-body systems can host a variety of unusual properties in their ground state, such as fractionalized quasiparticles and protected degeneracies.  In contrast, highly-excited states were long thought to be relatively boring, on the grounds that they should behave like thermal states, as dictated by the Eigenstate Thermalization Hypothesis (ETH)\cite{Deutsch,Rigol,Srednicki}.  In recent years, however, new types of quantum many-body systems have been studied which violate the ETH.  The most common example is many-body localization (MBL)\cite{GMP,BAA,mblarcmp}, typically driven by the effects of disorder, in which essentially all eigenstates are athermal, characterized by an extensive number of emergent local integrals of motion.

Recently, a new type of non-ergodic behavior has been observed in the form of quantum many-body scars\cite{turner2018weak, Serbyn2, LinMotrunich, Burnell}.  In contrast to the fully localized spectrum of MBL systems, scars are a small number of localized states in an otherwise thermalizing spectrum.  While scars constitute a vanishing fraction of the spectrum in the thermodynamic limit, they are of direct experimental relevance, since they have high overlap with easily-prepared product states.  Indeed, scar states have been proposed as an explanation for the long-time oscillations observed in Rydberg atom chains\cite{turner2018weak,lgt}.  The scar phenomenon, first encountered in the AKLT model \cite{moudgalya2018entanglement}, arises in a variety of many-body Hamiltonians \cite{ShiraishiMori, ChoiAbanin, ok2019topological}.  However, the origin and stability of athermal behavior in these models is not always intuitively clear.  To build a more systematic understanding of the scar phenomenon, it is desirable to identify mechanisms which give rise to scars on general grounds, independent of microscopic details.  In this paper, we demonstrate a fundamentally new type of scars, robust against arbitrary driving, which we refer to as ``dynamical scar states," arising in systems subject to certain conservation laws.  Specifically, we show how a small number of athermal states manifest in the steady-state configurations of a Floquet system as a consequence of a new mechanism for localization encountered in the context of fracton physics.

A fracton\cite{fractonarcmp} is an emergent quasiparticle found in various condensed matter contexts, such as spin liquids\cite{chamon,haah,fracton1,fracton2,slagle1, sondhi} and crystalline defects\cite{leomichael,pai, Gromov,pretko2018symmetry,kumar2018symmetry}, exhibiting a characteristic immobility arising from conservation of higher moments, such as dipole moment\cite{sub,genem}.  This constraint inhibits thermalization, since a fracton cannot freely move around the system.  In three spatial dimensions, a system of fractons will eventually thermalize, albeit logarithmically slowly, in a manifestation of glassy dynamics\cite{chamon, SivaYoshida, prem}. In one-dimensional fracton systems, however, a fracton can forever remain localized at its initial position, even under random local unitary time evolution.\cite{pai2018localization}  Unlike conventional localization, where particles are independently localized, a collection of fractons will agglomerate into a single peak at their center of mass, as a consequence of their gravitational attraction.\cite{mach}  Notably, only states featuring nonzero fracton charge can remain localized, while dipole states quickly thermalize.

The localization observed in random unitary circuits is expected to also manifest in the steady-state dynamics of Floquet fracton systems, which feature the extra constraint of conservation of quasienergy.  However, since fracton states can be localized while dipole states thermalize, it is clear that such a system cannot be fully localized.  Rather, we expect to see a special set of athermal states, as in the framework of many-body scars.  To consider the connection between fractons and scars in detail, we study a one-dimensional Floquet system with the mobility restrictions of fractons, implemented via quantum circuits.  In addition to the charge and dipole conservation characteristic of fracton systems, we further translation invariance, to rule out the possibility of conventional disorder-driven localization, but otherwise allow the unitary gates to be chosen randomly.

To determine the steady-state dynamics of this Floquet fracton system, we begin by finding the spectrum of the Floquet operator, which contains many ETH-violating eigenstates, as discussed in the context of Hilbert space ``fragmentation".\cite{ah1,ah2}  We then consider a more generic basis of initial conditions which are not eigenstates of the Floquet evolution.  For a typical basis of initial conditions, the majority of states heat up to an entropy-maximizing thermal state at near-infinite temperature.  In contrast, a small number of states remain stably localized under the driving, characterized by subthermal entanglement.  We refer to these athermal steady states as ``dynamical scar states," in analogy with the scar eigenstates of Hamiltonian systems.  There is one scar state in each sector of a particular charge and dipole moment, characterized by an agglomerated fracton peak.  The number of scar states grows algebraically with system size, $L^3$, while the number of thermalizing states grows exponentially, $3^L$.  While the scar states represent a vanishing fraction of the spectrum in the thermodynamic limit, they are experimentally relevant due to their high overlap with product states.  These localized states appear as a direct consequence of conservation of dipole moment.\cite{pai2018localization}  This opens the possibility that other types of scar states may occur as a consequence of some similarly simple physical principle, without depending on microscopic details.

\begin{figure}[t]
\captionsetup{singlelinecheck = false, justification=raggedright}
\centering
 \includegraphics[scale=0.38]{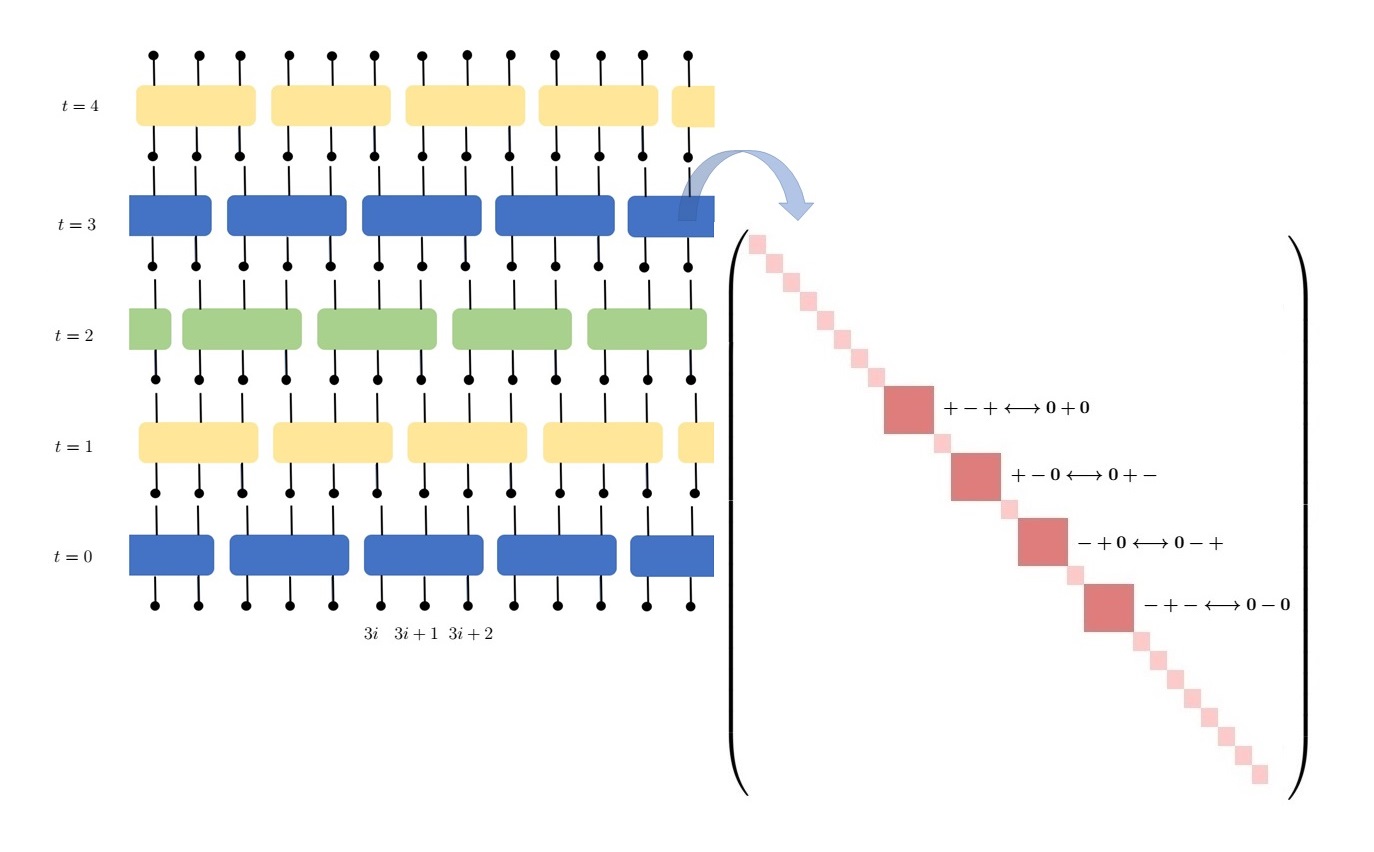}
 \caption{Floquet random unitary circuit (period 3): each site is a three-state qudit. Each gate (colored box) conserves $S^{total}_{z}$ and $\vec{P}_{total}$ of the three qudits it acts upon. The block diagonal Haar-random unitary with its nontrivial blocks is also shown.  All gates of a particular color are identical (to ensure translation-invariance).}
 \label{fig:floquetCircuit}
 \end{figure}

\emph{Fractonic Floquet Quantum Circuit Model}.    We work with a one-dimensional chain of $L$ sites, with a single spin-1 on each site, and periodic boundary conditions.  We time-evolve with a random quantum circuit of local unitary gates, constrained to locally conserve the total $z$ component of the spins (which serves as a conserved $U(1)$ charge), and also the total dipole moment of this effective charge (evaluated with respect to an arbitrary origin, and conserved mod $L$ due to periodic boundary conditions). Instead of a completely random unitary circuit, as in Ref. \onlinecite{pai2018localization}, we impose discrete time-translation symmetry. Using a stroboscopically repeating circuit allows us to study eigenstates and eigenvalues, \textit{i.e.} providing more tools compared to a simple random circuit. We consider a translation-invariant Floquet random circuit (see Figure \ref{fig:floquetCircuit}) to exclude the possibility of localization for conventional reasons.  The time-evolution is governed by a circuit with staggered layers of three-site unitary gates. The time evolution unitary is given by 
$U(t) = \prod_{t'=1}^{t}U(t',t'-1)$, where 
\begin{equation}
  U(t',t'-1)=\begin{cases}
    \prod_{i}U^{A}_{3i,3i+1,3i+2} & \text{if $t'$ mod $3 = 0$}\\
    \prod_{i}U^{B}_{3i-1,3i,3i+1} & \text{if $t'$ mod $3 = 1$}\\
    \prod_{i}U^{C}_{3i-2,3i-1,3i} & \text{if $t'$ mod $3 = 2$},
  \end{cases}
\end{equation}
and $U^{A}, U^{B}$, and $U^{C}$ are chosen at random for a given realization, but remain fixed throughout that run.

\begin{figure*}[t]
\captionsetup{singlelinecheck = false, justification=raggedright}
\centering
 \includegraphics[scale=0.56]{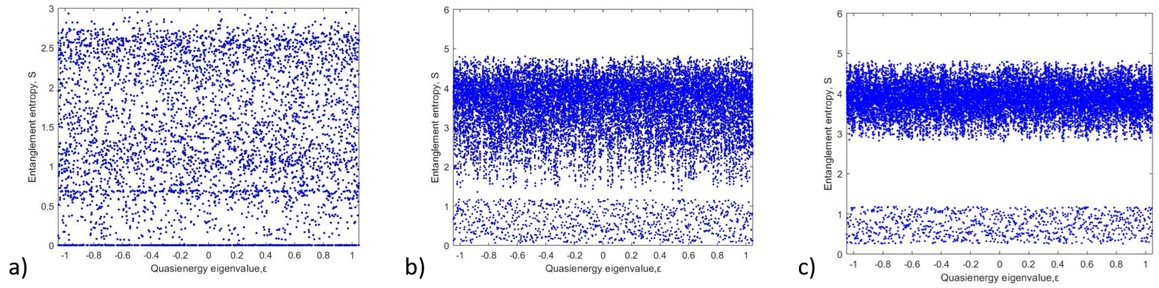}
 \caption{Bipartite entropy $S$ of: \textbf{a)} pure eigenstates, exhibiting Hilbert space fragmentation\cite{ah1,ah2}, \textbf{b}) steady states of initial conditions slightly different from eigenstates ($\Delta\varepsilon = 10^{-3}$), and \textbf{c}) steady states of random initial conditions.}
 \label{fig:SvsE}
 \end{figure*}

\emph{Steady States}.    To study the dynamics of our Floquet fractonic circuit, we begin by finding the eigenstates of the Floquet operator, which are trivially steady states.  If the system is initialized in any of these eigenstates, then it will remain in that state for all later times.  These eigenstates can be characterized in terms of their entanglement.  In Figure \ref{fig:SvsE}a, we plot the bipartite entanglement entropy of these eigenstates as a function of their quasienergy (over $[-\pi/3,\pi/3]$).  In contrast to a simple thermalizing system, in which most eigenstates have near-maximal thermal entanglement, the eigenstates of our Floquet fractonic circuit have a wide range of entanglement values.  In particular, there are a number of zero-entanglement product states in the spectrum, arising as a consequence of the Hilbert space ``fragmentation" discussed in Refs. \onlinecite{ah1,ah2}.

While the eigenstates $|\psi_n\rangle$ of the Floquet operator exhibit a large degree of athermal behavior, it is important to consider a more general set of initial conditions for our fractonic circuit.  Say we prepare the system in a state from a different basis, $|\phi_m\rangle$, prior to applying a Floquet fractonic circuit.  For example, we could prepare the system in an eigenstate of some other Hamiltonian.  After time-evolving by time $t$, the state of the system will be:
\begin{equation}
|\Phi_m(t)\rangle = \sum_n e^{i\epsilon_n t}|\psi_n\rangle \langle \psi_n|\phi_m\rangle
\end{equation}
where $\epsilon_n$ is the quasienergy of eigenstate $|\psi_n\rangle$.  We now form the density matrix, $\rho_m(t) = |\Phi_m(t)\rangle\langle\Phi_m(t)|$, and take its time average to find the steady state of the Floquet time evolution.  Assuming negligible degeneracies in the spectrum (as borne out by the data in Figure \ref{fig:SvsE}a), the steady state of the system is given by\cite{chandran}:
\begin{equation}
\overline{\rho}_m = \sum_n |A_{mn}|^2|\psi_n\rangle\langle \psi_n|
\end{equation}
where $A_{mn} = \langle\psi_n|\phi_m\rangle$.  We therefore see that we can form steady states of the Floquet fractonic circuit by simply taking linear combinations of the density matrices of the eigenstates.

We first investigate a set of initial conditions which are only mildly changed from the eigenstate basis.  We consider initializing our system in states $|\phi_m\rangle$ which are random superpositions of eigenstates only within some quasienergy window $\Delta \varepsilon$, such that eigenstates are recovered in the $\Delta \varepsilon\rightarrow 0$ limit.  In Figure \ref{fig:SvsE}b, we plot the bipartite entropy ($i.e.$ the entropy of the reduced density matrix for half of the system) of the steady states versus their average quasienergy, for $\Delta \varepsilon = 10^{-3}$.  Even for this small deviation from eigenstates, the states begin to separate into two distinct entropy bands, unlike the seemingly random entropies of eigenstates.  The majority of states exist in a band near maximal entropy, consistent with an infinite temperature state.  Meanwhile, a much smaller set of states exhibit significantly lower entanglement.  These athermal steady states have average quasienergies scattered fairly evenly throughout the quasienergy spectrum.

To confirm the generality of this picture for typical initial conditions, we next consider a randomly chosen basis $|\phi_m\rangle$ of initial conditions.  In other words, we let the energy window $\Delta\varepsilon$ of superpositions tend to $2\pi/3$.  The bipartite entropy of the steady states versus their average quasienergy is plotted in Figure \ref{fig:SvsE}c.  As can be seen, a randomly chosen basis of initial conditions leads to two fairly sharp entropy bands, with the lower band having a clearly subthermal entropy.  The existence of these low-entropy states provides a counter-example to the conventional wisdom that a Floquet system should always heat up to infinite temperature unless its spectrum is completely localized.  In contrast, our Floquet fractonic circuit only fails to thermalize for certain special initial conditions.

Importantly, the number of low-entanglement states grows only algebraically in system size, as we discuss below, while the number of thermal states grows exponentially.  In light of these facts, we refer to these steady states as ``dynamical scar states," in analogy with the ETH-violating scar eigenstates of certain Hamiltonian systems.  The existence of these scar steady states under Floquet evolution is independent of the details of the gates making up the time evolution evolution operator, which are chosen randomly.  Furthermore, the dynamical scar states are present even for a translation-invariant Floquet random circuit, indicating that scarring does not arise from conventional disorder-driven localization.  This is consistent with the behavior of fully random fractonic circuits, which were similarly argued to exhibit localization in the absence of disorder.\cite{pai2018localization}

\begin{figure*}[t]
\captionsetup{singlelinecheck = false, justification=raggedright}
\centering
 \includegraphics[scale=0.54]{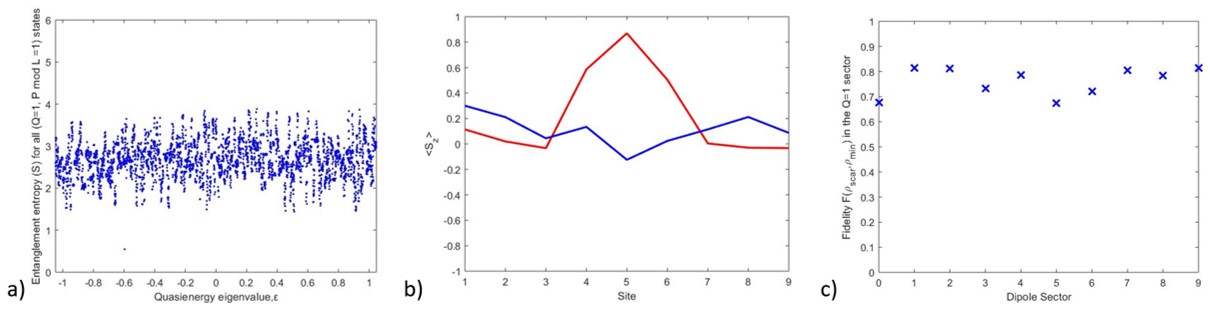}
 \caption{\textbf{a}) For typical initial conditions, there is one scar state per sector, as diagnosed by entropy (shown for $Q=1,P=1$).  \textbf{b}) Scar states (red) feature an agglomerated fracton peak, while thermal states (blue) have a mostly flat $\langle S_z\rangle$ profile.  \textbf{c}) Scar states have high fidelity with minimal product density matrices.  (All data for $L=9$.)}
 \label{fig:sector}
 \end{figure*}

\emph{Characterization of Scar States}.    To build intuition for the nature of the dynamical scar states, it is useful to study the profile of the $S_z$ expectation value.  In Figure \ref{fig:sector}b, we display the $\langle S_z\rangle$ profile for a typical scar state and typical thermal state as a function of position for an $L=9$ system.  The thermal state has an almost flat distribution, as expected.  In contrast, the scar states each feature a single localized fracton peak.  Even for initial conditions with multiple fractons scattered throughout the system, the steady-state configuration features only a single peak, corresponding to the fractons clustering at their mutual center of mass.  For each $(Q,P)$ sector, there is only a single scar steady state (see Figure \ref{fig:sector}a) with the fractons maximally clustered.  The only exception is the $Q=0$ sector, which does not exhibit any localized states.  This behavior is consistent with the fracton agglomeration observed in the steady states of fractonic random circuits.\cite{pai2018localization}

Remarkably, the scar states have high overlap with ``minimal" product density matrices corresponding to different values of charge and dipole moment.  The minimal product density matrix with charge $Q$ and dipole moment $P$ is a product of identity operators on almost every site, except for $(I+S^z)$ operators on exactly $Q$ sites chosen to correspond to dipole moment $P$.  For example, for $Q=1$, the corresponding minimal product density matrices takes the form $\rho_{\textrm{min}} = \cdot\cdot\cdot I\otimes I\otimes (I+S_z)\otimes I\otimes I \cdot\cdot$, where the lone $(I+S_z)$ operator is on site $P$ (with respect to the chosen origin).  For higher charges, the $(I+S_z)$ operators are placed as close together as possible consistent with the given dipole moment, to capture the effects of fracton agglomeration.  We now evaluate the quantum fidelity between a scar steady state, $\rho_{\textrm{scar}}$, and the minimal product density matrix with the same $Q$ and $P$ expectation values:
\begin{equation}
F(\rho_{\textrm{min}},\rho_{\textrm{scar}}) = \bigg( \textrm{Tr}\bigg[\sqrt{\sqrt{\rho_{\textrm{min}}}\rho_{\textrm{scar}}\sqrt{\rho_{\textrm{min}}}}\bigg]  \bigg)^2
\end{equation}
which serves as a measure of closeness of the two quantum states.  The results are shown in Figure \ref{fig:sector}c for states in the $Q=1$ sector of an $L=9$ system.  We find high agreement between the scar states and minimal product density matrices.  These minimal product density matrices were precisely the initial conditions which led to fracton localization under random unitary circuit dynamics.\cite{pai2018localization}  We therefore identify scars with the localized steady states observed in Ref. \onlinecite{pai2018localization}, and conclude that scarring originates from the same physical mechanism.

\begin{figure*}[t]
\captionsetup{singlelinecheck = false, justification=raggedright}
\centering
 \includegraphics[scale=0.56]{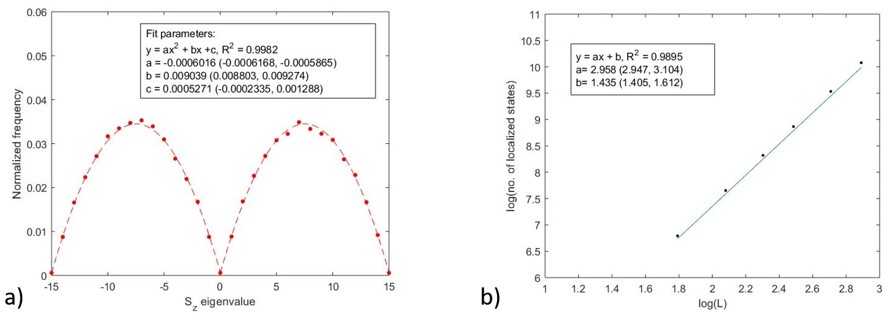}
 \caption{\textbf{a}) Frequency (normalized by total number of scar states) of the number of dipole sectors per $S_{z}$ eigenvalue $Q$. Dashed line is analytic prediction, dots are data $(L=15)$.  \textbf{b}) The number of scar states scale as $\sim L^{3}$, while the total number of states scales exponentially, \textit{i.e.} $\sim 3^{L}$.}
 \label{fig:enum}
 \end{figure*}

Our present investigation thus suggests that quantum dynamics with fractonic constraints is ergodic almost everywhere in Hilbert space, as characterized by near-maximal entropy.  However, there is a special scar subregion of Hilbert space that displays nonergodic behavior under driving.  Since these localized states have high overlap with the minimal product density matrix initial conditions considered in Ref. \onlinecite{pai2018localization}, we can understand the ergodicity breaking in the scar regions in terms of the fracton localization mechanism discussed therein.  Furthermore, being close to product states, the scar states are of direct experimental importance.

\emph{Enumerating the Scar States}.    As we have seen earlier, for typical initial conditions, there is precisely one scar steady state per sector ($Q,P$), corresponding to the minimal product density matrix within that sector. Therefore, to determine the number of scar steady states for a given basis, we only need to count the distinct number of $(Q,P)$ sectors.  We first determine the number of distinct dipole sectors for a given charge $Q$. For a system of size $L$, given a value of charge $Q$, the value of the dipole moment can go from $Q(Q-1)/2$ to $QL-Q(Q+1)/2$. This gives $QL-Q^{2} + 1$ distinct sectors per charge $Q$. Note that this formula is not operative for the zero-fracton ($Q=0$) sector, where there is no localization and we therefore expect no scar states. This formula agrees well with what we observe in our simulations (see Figure \ref{fig:enum}a). 

Now we determine the number of scar steady states $\mathcal{N}^{\textrm{total}}_{\textrm{scar}}(L)$ in the entire spectrum for a system of size $L$, and test our analytic prediction against numerics. To do this, we evaluate the following sum:
\begin{equation}
\mathcal{N}^{\textrm{total}}_{\textrm{scar}}(L) = 2\sum_{Q=1}^{L}(QL - Q^{2}+1).
\end{equation}
This sum gives $\mathcal{N}^{\textrm{total}}_{\textrm{scar}}(L)= L^{3}/3 + 5L/3$ \textit{i.e.} $\mathcal{N}^{\textrm{total}}_{\textrm{scar}}(L) \sim L^{3}$. We verify this scaling numerically in Figure \ref{fig:enum}b. The good agreement between the counting of minimal product density matrices and the observed number of scar states gives us additional confidence in our interpretation.  Note that the scar states constitute only a tiny fraction of the total Hilbert space, which has $3^L$ states, most of which are thermal.

\emph{Discussion and Conclusions}.    In this work, we have shown how the conservation laws associated with fracton systems, such as conservation of charge and dipole moment, lead to athermal behavior in the steady states of a Floquet system.  Specifically, for a typical basis of initial conditions, athermality is manifested in a small set of states which remain localized under the driving, while the majority of initial conditions heat up to an infinite-temperature steady state.  We refer to this new type of athermal state as a ``dynamical scar state," in analogy with the scar eigenstates observed in Hamiltonian systems.  These scar states represent a vanishingly small fraction of the total Hilbert space in the thermodynamic limit, but are nevertheless experimentally relevant due to their high overlap with easily prepared product states.  The scar states correspond to agglomerated fracton peaks, which are the expected late-time configurations associated with fracton localization.  Fracton systems therefore provide a novel manifestation of many-body scars which do not depend on microscopic details, but rather follow from a simple physical principle, namely the higher moment conservation laws associated with fractons.  We hope that this analysis may yield more general insights about the physical mechanism behind quantum many-body scars.

\begin{acknowledgments}
We thank Rahul Nandkishore for useful discussions and a prior collaboration.  We also thank David Huse, Vedika Khemani, Pablo Sala, Tibor Rakovszky, Ruben Verresen, Michael Knap, and Frank Pollmann for discussions and feedback on the manuscript.  This material is based upon work supported by the Air Force Office of Scientific Research under award number FA9550-17-1-0183.
\end{acknowledgments}

\bibliography{library}

\begin{thebibliography}{36}%
\makeatletter
\providecommand \@ifxundefined [1]{%
 \@ifx{#1\undefined}
}%
\providecommand \@ifnum [1]{%
 \ifnum #1\expandafter \@firstoftwo
 \else \expandafter \@secondoftwo
 \fi
}%
\providecommand \@ifx [1]{%
 \ifx #1\expandafter \@firstoftwo
 \else \expandafter \@secondoftwo
 \fi
}%
\providecommand \natexlab [1]{#1}%
\providecommand \enquote  [1]{``#1''}%
\providecommand \bibnamefont  [1]{#1}%
\providecommand \bibfnamefont [1]{#1}%
\providecommand \citenamefont [1]{#1}%
\providecommand \href@noop [0]{\@secondoftwo}%
\providecommand \href [0]{\begingroup \@sanitize@url \@href}%
\providecommand \@href[1]{\@@startlink{#1}\@@href}%
\providecommand \@@href[1]{\endgroup#1\@@endlink}%
\providecommand \@sanitize@url [0]{\catcode `\\12\catcode `\$12\catcode
  `\&12\catcode `\#12\catcode `\^12\catcode `\_12\catcode `\%12\relax}%
\providecommand \@@startlink[1]{}%
\providecommand \@@endlink[0]{}%
\providecommand \url  [0]{\begingroup\@sanitize@url \@url }%
\providecommand \@url [1]{\endgroup\@href {#1}{\urlprefix }}%
\providecommand \urlprefix  [0]{URL }%
\providecommand \Eprint [0]{\href }%
\providecommand \doibase [0]{http://dx.doi.org/}%
\providecommand \selectlanguage [0]{\@gobble}%
\providecommand \bibinfo  [0]{\@secondoftwo}%
\providecommand \bibfield  [0]{\@secondoftwo}%
\providecommand \translation [1]{[#1]}%
\providecommand \BibitemOpen [0]{}%
\providecommand \bibitemStop [0]{}%
\providecommand \bibitemNoStop [0]{.\EOS\space}%
\providecommand \EOS [0]{\spacefactor3000\relax}%
\providecommand \BibitemShut  [1]{\csname bibitem#1\endcsname}%
\let\auto@bib@innerbib\@empty
\bibitem [{\citenamefont {Deutsch}(1991)}]{Deutsch}%
  \BibitemOpen
  \bibfield  {author} {\bibinfo {author} {\bibfnamefont {J.~M.}\ \bibnamefont
  {Deutsch}},\ }\href {\doibase 10.1103/PhysRevA.43.2046} {\bibfield  {journal}
  {\bibinfo  {journal} {Phys. Rev. A}\ }\textbf {\bibinfo {volume} {43}},\
  \bibinfo {pages} {2046} (\bibinfo {year} {1991})}\BibitemShut {NoStop}%
\bibitem [{\citenamefont {Rigol}\ \emph {et~al.}(2008)\citenamefont {Rigol},
  \citenamefont {Dunjko},\ and\ \citenamefont {Olshanii}}]{Rigol}%
  \BibitemOpen
  \bibfield  {author} {\bibinfo {author} {\bibfnamefont {M.}~\bibnamefont
  {Rigol}}, \bibinfo {author} {\bibfnamefont {V.}~\bibnamefont {Dunjko}}, \
  and\ \bibinfo {author} {\bibfnamefont {M.}~\bibnamefont {Olshanii}},\ }\href
  {http://dx.doi.org/10.1038/nature06838} {\bibfield  {journal} {\bibinfo
  {journal} {Nature}\ }\textbf {\bibinfo {volume} {452}},\ \bibinfo {pages}
  {854} (\bibinfo {year} {2008})}\BibitemShut {NoStop}%
\bibitem [{\citenamefont {Srednicki}(1994)}]{Srednicki}%
  \BibitemOpen
  \bibfield  {author} {\bibinfo {author} {\bibfnamefont {M.}~\bibnamefont
  {Srednicki}},\ }\href {\doibase 10.1103/PhysRevE.50.888} {\bibfield
  {journal} {\bibinfo  {journal} {Phys. Rev. E}\ }\textbf {\bibinfo {volume}
  {50}},\ \bibinfo {pages} {888} (\bibinfo {year} {1994})}\BibitemShut
  {NoStop}%
\bibitem [{\citenamefont {Gornyi}\ \emph {et~al.}(2005)\citenamefont {Gornyi},
  \citenamefont {Mirlin},\ and\ \citenamefont {Polyakov}}]{GMP}%
  \BibitemOpen
  \bibfield  {author} {\bibinfo {author} {\bibfnamefont {I.~V.}\ \bibnamefont
  {Gornyi}}, \bibinfo {author} {\bibfnamefont {A.~D.}\ \bibnamefont {Mirlin}},
  \ and\ \bibinfo {author} {\bibfnamefont {D.~G.}\ \bibnamefont {Polyakov}},\
  }\href {\doibase 10.1103/PhysRevLett.95.206603} {\bibfield  {journal}
  {\bibinfo  {journal} {Phys. Rev. Lett.}\ }\textbf {\bibinfo {volume} {95}},\
  \bibinfo {pages} {206603} (\bibinfo {year} {2005})}\BibitemShut {NoStop}%
\bibitem [{\citenamefont {Basko}\ \emph {et~al.}(2006)\citenamefont {Basko},
  \citenamefont {Aleiner},\ and\ \citenamefont {Altshuler}}]{BAA}%
  \BibitemOpen
  \bibfield  {author} {\bibinfo {author} {\bibfnamefont {D.}~\bibnamefont
  {Basko}}, \bibinfo {author} {\bibfnamefont {I.}~\bibnamefont {Aleiner}}, \
  and\ \bibinfo {author} {\bibfnamefont {B.}~\bibnamefont {Altshuler}},\ }\href
  {\doibase https://doi.org/10.1016/j.aop.2005.11.014} {\bibfield  {journal}
  {\bibinfo  {journal} {Annals of Physics}\ }\textbf {\bibinfo {volume}
  {321}},\ \bibinfo {pages} {1126 } (\bibinfo {year} {2006})}\BibitemShut
  {NoStop}%
\bibitem [{\citenamefont {Nandkishore}\ and\ \citenamefont
  {Huse}(2015)}]{mblarcmp}%
  \BibitemOpen
  \bibfield  {author} {\bibinfo {author} {\bibfnamefont {R.}~\bibnamefont
  {Nandkishore}}\ and\ \bibinfo {author} {\bibfnamefont {D.~A.}\ \bibnamefont
  {Huse}},\ }\href {\doibase 10.1146/annurev-conmatphys-031214-014726}
  {\bibfield  {journal} {\bibinfo  {journal} {Annual Review of Condensed Matter
  Physics}\ }\textbf {\bibinfo {volume} {6}},\ \bibinfo {pages} {15} (\bibinfo
  {year} {2015})},\ \Eprint
  {http://arxiv.org/abs/https://doi.org/10.1146/annurev-conmatphys-031214-014726}
  {https://doi.org/10.1146/annurev-conmatphys-031214-014726} \BibitemShut
  {NoStop}%
\bibitem [{\citenamefont {Turner}\ \emph
  {et~al.}(2018{\natexlab{a}})\citenamefont {Turner}, \citenamefont
  {Michailidis}, \citenamefont {Abanin}, \citenamefont {Serbyn},\ and\
  \citenamefont {Papi{\'c}}}]{turner2018weak}%
  \BibitemOpen
  \bibfield  {author} {\bibinfo {author} {\bibfnamefont {C.}~\bibnamefont
  {Turner}}, \bibinfo {author} {\bibfnamefont {A.}~\bibnamefont {Michailidis}},
  \bibinfo {author} {\bibfnamefont {D.}~\bibnamefont {Abanin}}, \bibinfo
  {author} {\bibfnamefont {M.}~\bibnamefont {Serbyn}}, \ and\ \bibinfo {author}
  {\bibfnamefont {Z.}~\bibnamefont {Papi{\'c}}},\ }\href@noop {} {\bibfield
  {journal} {\bibinfo  {journal} {Nature Physics}\ } (\bibinfo {year}
  {2018}{\natexlab{a}})}\BibitemShut {NoStop}%
\bibitem [{\citenamefont {Turner}\ \emph
  {et~al.}(2018{\natexlab{b}})\citenamefont {Turner}, \citenamefont
  {Michailidis}, \citenamefont {Abanin}, \citenamefont {Serbyn},\ and\
  \citenamefont {Papi\ifmmode~\acute{c}\else \'{c}\fi{}}}]{Serbyn2}%
  \BibitemOpen
  \bibfield  {author} {\bibinfo {author} {\bibfnamefont {C.~J.}\ \bibnamefont
  {Turner}}, \bibinfo {author} {\bibfnamefont {A.~A.}\ \bibnamefont
  {Michailidis}}, \bibinfo {author} {\bibfnamefont {D.~A.}\ \bibnamefont
  {Abanin}}, \bibinfo {author} {\bibfnamefont {M.}~\bibnamefont {Serbyn}}, \
  and\ \bibinfo {author} {\bibfnamefont {Z.}~\bibnamefont
  {Papi\ifmmode~\acute{c}\else \'{c}\fi{}}},\ }\href {\doibase
  10.1103/PhysRevB.98.155134} {\bibfield  {journal} {\bibinfo  {journal} {Phys.
  Rev. B}\ }\textbf {\bibinfo {volume} {98}},\ \bibinfo {pages} {155134}
  (\bibinfo {year} {2018}{\natexlab{b}})}\BibitemShut {NoStop}%
\bibitem [{\citenamefont {{Lin}}\ and\ \citenamefont
  {{Motrunich}}(2018)}]{LinMotrunich}%
  \BibitemOpen
  \bibfield  {author} {\bibinfo {author} {\bibfnamefont {C.-J.}\ \bibnamefont
  {{Lin}}}\ and\ \bibinfo {author} {\bibfnamefont {O.~I.}\ \bibnamefont
  {{Motrunich}}},\ }\href@noop {} {\bibfield  {journal} {\bibinfo  {journal}
  {arXiv e-prints}\ ,\ \bibinfo {eid} {arXiv:1810.00888}} (\bibinfo {year}
  {2018})},\ \Eprint {http://arxiv.org/abs/1810.00888} {arXiv:1810.00888
  [cond-mat.quant-gas]} \BibitemShut {NoStop}%
\bibitem [{\citenamefont {Chen}\ \emph {et~al.}(2018)\citenamefont {Chen},
  \citenamefont {Burnell},\ and\ \citenamefont {Chandran}}]{Burnell}%
  \BibitemOpen
  \bibfield  {author} {\bibinfo {author} {\bibfnamefont {C.}~\bibnamefont
  {Chen}}, \bibinfo {author} {\bibfnamefont {F.}~\bibnamefont {Burnell}}, \
  and\ \bibinfo {author} {\bibfnamefont {A.}~\bibnamefont {Chandran}},\ }\href
  {\doibase 10.1103/PhysRevLett.121.085701} {\bibfield  {journal} {\bibinfo
  {journal} {Phys. Rev. Lett.}\ }\textbf {\bibinfo {volume} {121}},\ \bibinfo
  {pages} {085701} (\bibinfo {year} {2018})}\BibitemShut {NoStop}%
\bibitem [{\citenamefont {{Surace}}\ \emph {et~al.}(2019)\citenamefont
  {{Surace}}, \citenamefont {{Mazza}}, \citenamefont {{Giudici}}, \citenamefont
  {{Lerose}}, \citenamefont {{Gambassi}},\ and\ \citenamefont
  {{Dalmonte}}}]{lgt}%
  \BibitemOpen
  \bibfield  {author} {\bibinfo {author} {\bibfnamefont {F.~M.}\ \bibnamefont
  {{Surace}}}, \bibinfo {author} {\bibfnamefont {P.~P.}\ \bibnamefont
  {{Mazza}}}, \bibinfo {author} {\bibfnamefont {G.}~\bibnamefont {{Giudici}}},
  \bibinfo {author} {\bibfnamefont {A.}~\bibnamefont {{Lerose}}}, \bibinfo
  {author} {\bibfnamefont {A.}~\bibnamefont {{Gambassi}}}, \ and\ \bibinfo
  {author} {\bibfnamefont {M.}~\bibnamefont {{Dalmonte}}},\ }\href@noop {}
  {\bibfield  {journal} {\bibinfo  {journal} {arXiv e-prints}\ ,\ \bibinfo
  {eid} {arXiv:1902.09551}} (\bibinfo {year} {2019})},\ \Eprint
  {http://arxiv.org/abs/1902.09551} {arXiv:1902.09551 [cond-mat.quant-gas]}
  \BibitemShut {NoStop}%
\bibitem [{\citenamefont {Moudgalya}\ \emph {et~al.}(2018)\citenamefont
  {Moudgalya}, \citenamefont {Regnault},\ and\ \citenamefont
  {Bernevig}}]{moudgalya2018entanglement}%
  \BibitemOpen
  \bibfield  {author} {\bibinfo {author} {\bibfnamefont {S.}~\bibnamefont
  {Moudgalya}}, \bibinfo {author} {\bibfnamefont {N.}~\bibnamefont {Regnault}},
  \ and\ \bibinfo {author} {\bibfnamefont {B.~A.}\ \bibnamefont {Bernevig}},\
  }\href@noop {} {\bibfield  {journal} {\bibinfo  {journal} {arXiv preprint
  arXiv:1806.09624}\ } (\bibinfo {year} {2018})}\BibitemShut {NoStop}%
\bibitem [{\citenamefont {Shiraishi}\ and\ \citenamefont
  {Mori}(2017)}]{ShiraishiMori}%
  \BibitemOpen
  \bibfield  {author} {\bibinfo {author} {\bibfnamefont {N.}~\bibnamefont
  {Shiraishi}}\ and\ \bibinfo {author} {\bibfnamefont {T.}~\bibnamefont
  {Mori}},\ }\href {\doibase 10.1103/PhysRevLett.119.030601} {\bibfield
  {journal} {\bibinfo  {journal} {Phys. Rev. Lett.}\ }\textbf {\bibinfo
  {volume} {119}},\ \bibinfo {pages} {030601} (\bibinfo {year}
  {2017})}\BibitemShut {NoStop}%
\bibitem [{\citenamefont {{Choi}}\ \emph {et~al.}(2018)\citenamefont {{Choi}},
  \citenamefont {{Turner}}, \citenamefont {{Pichler}}, \citenamefont {{Ho}},
  \citenamefont {{Michailidis}}, \citenamefont {{Papi{\'c}}}, \citenamefont
  {{Serbyn}}, \citenamefont {{Lukin}},\ and\ \citenamefont
  {{Abanin}}}]{ChoiAbanin}%
  \BibitemOpen
  \bibfield  {author} {\bibinfo {author} {\bibfnamefont {S.}~\bibnamefont
  {{Choi}}}, \bibinfo {author} {\bibfnamefont {C.~J.}\ \bibnamefont
  {{Turner}}}, \bibinfo {author} {\bibfnamefont {H.}~\bibnamefont {{Pichler}}},
  \bibinfo {author} {\bibfnamefont {W.~W.}\ \bibnamefont {{Ho}}}, \bibinfo
  {author} {\bibfnamefont {A.~A.}\ \bibnamefont {{Michailidis}}}, \bibinfo
  {author} {\bibfnamefont {Z.}~\bibnamefont {{Papi{\'c}}}}, \bibinfo {author}
  {\bibfnamefont {M.}~\bibnamefont {{Serbyn}}}, \bibinfo {author}
  {\bibfnamefont {M.~D.}\ \bibnamefont {{Lukin}}}, \ and\ \bibinfo {author}
  {\bibfnamefont {D.~A.}\ \bibnamefont {{Abanin}}},\ }\href@noop {} {\bibfield
  {journal} {\bibinfo  {journal} {arXiv e-prints}\ ,\ \bibinfo {eid}
  {arXiv:1812.05561}} (\bibinfo {year} {2018})},\ \Eprint
  {http://arxiv.org/abs/1812.05561} {arXiv:1812.05561 [quant-ph]} \BibitemShut
  {NoStop}%
\bibitem [{\citenamefont {Ok}\ \emph {et~al.}(2019)\citenamefont {Ok},
  \citenamefont {Choo}, \citenamefont {Mudry}, \citenamefont {Castelnovo},
  \citenamefont {Chamon},\ and\ \citenamefont {Neupert}}]{ok2019topological}%
  \BibitemOpen
  \bibfield  {author} {\bibinfo {author} {\bibfnamefont {S.}~\bibnamefont
  {Ok}}, \bibinfo {author} {\bibfnamefont {K.}~\bibnamefont {Choo}}, \bibinfo
  {author} {\bibfnamefont {C.}~\bibnamefont {Mudry}}, \bibinfo {author}
  {\bibfnamefont {C.}~\bibnamefont {Castelnovo}}, \bibinfo {author}
  {\bibfnamefont {C.}~\bibnamefont {Chamon}}, \ and\ \bibinfo {author}
  {\bibfnamefont {T.}~\bibnamefont {Neupert}},\ }\href@noop {} {\bibfield
  {journal} {\bibinfo  {journal} {arXiv preprint arXiv:1901.01260}\ } (\bibinfo
  {year} {2019})}\BibitemShut {NoStop}%
\bibitem [{\citenamefont {{Nandkishore}}\ and\ \citenamefont
  {{Hermele}}(2018)}]{fractonarcmp}%
  \BibitemOpen
  \bibfield  {author} {\bibinfo {author} {\bibfnamefont {R.~M.}\ \bibnamefont
  {{Nandkishore}}}\ and\ \bibinfo {author} {\bibfnamefont {M.}~\bibnamefont
  {{Hermele}}},\ }\href@noop {} {\bibfield  {journal} {\bibinfo  {journal}
  {ArXiv e-prints}\ } (\bibinfo {year} {2018})},\ \Eprint
  {http://arxiv.org/abs/1803.11196} {arXiv:1803.11196 [cond-mat.str-el]}
  \BibitemShut {NoStop}%
\bibitem [{\citenamefont {Chamon}(2005)}]{chamon}%
  \BibitemOpen
  \bibfield  {author} {\bibinfo {author} {\bibfnamefont {C.}~\bibnamefont
  {Chamon}},\ }\href {\doibase 10.1103/PhysRevLett.94.040402} {\bibfield
  {journal} {\bibinfo  {journal} {Phys. Rev. Lett.}\ }\textbf {\bibinfo
  {volume} {94}},\ \bibinfo {pages} {040402} (\bibinfo {year}
  {2005})}\BibitemShut {NoStop}%
\bibitem [{\citenamefont {Haah}(2011)}]{haah}%
  \BibitemOpen
  \bibfield  {author} {\bibinfo {author} {\bibfnamefont {J.}~\bibnamefont
  {Haah}},\ }\href {\doibase 10.1103/PhysRevA.83.042330} {\bibfield  {journal}
  {\bibinfo  {journal} {Phys. Rev. A}\ }\textbf {\bibinfo {volume} {83}},\
  \bibinfo {pages} {042330} (\bibinfo {year} {2011})}\BibitemShut {NoStop}%
\bibitem [{\citenamefont {Vijay}\ \emph {et~al.}(2015)\citenamefont {Vijay},
  \citenamefont {Haah},\ and\ \citenamefont {Fu}}]{fracton1}%
  \BibitemOpen
  \bibfield  {author} {\bibinfo {author} {\bibfnamefont {S.}~\bibnamefont
  {Vijay}}, \bibinfo {author} {\bibfnamefont {J.}~\bibnamefont {Haah}}, \ and\
  \bibinfo {author} {\bibfnamefont {L.}~\bibnamefont {Fu}},\ }\href {\doibase
  10.1103/PhysRevB.92.235136} {\bibfield  {journal} {\bibinfo  {journal} {Phys.
  Rev. B}\ }\textbf {\bibinfo {volume} {92}},\ \bibinfo {pages} {235136}
  (\bibinfo {year} {2015})}\BibitemShut {NoStop}%
\bibitem [{\citenamefont {Vijay}\ \emph {et~al.}(2016)\citenamefont {Vijay},
  \citenamefont {Haah},\ and\ \citenamefont {Fu}}]{fracton2}%
  \BibitemOpen
  \bibfield  {author} {\bibinfo {author} {\bibfnamefont {S.}~\bibnamefont
  {Vijay}}, \bibinfo {author} {\bibfnamefont {J.}~\bibnamefont {Haah}}, \ and\
  \bibinfo {author} {\bibfnamefont {L.}~\bibnamefont {Fu}},\ }\href {\doibase
  10.1103/PhysRevB.94.235157} {\bibfield  {journal} {\bibinfo  {journal} {Phys.
  Rev. B}\ }\textbf {\bibinfo {volume} {94}},\ \bibinfo {pages} {235157}
  (\bibinfo {year} {2016})}\BibitemShut {NoStop}%
\bibitem [{\citenamefont {{Slagle}}\ and\ \citenamefont
  {{Kim}}(2017)}]{slagle1}%
  \BibitemOpen
  \bibfield  {author} {\bibinfo {author} {\bibfnamefont {K.}~\bibnamefont
  {{Slagle}}}\ and\ \bibinfo {author} {\bibfnamefont {Y.~B.}\ \bibnamefont
  {{Kim}}},\ }\href@noop {} {\bibfield  {journal} {\bibinfo  {journal} {ArXiv
  e-prints}\ } (\bibinfo {year} {2017})},\ \Eprint
  {http://arxiv.org/abs/1704.03870} {arXiv:1704.03870} \BibitemShut {NoStop}%
\bibitem [{\citenamefont {{You}}\ \emph {et~al.}(2018)\citenamefont {{You}},
  \citenamefont {{Devakul}}, \citenamefont {{Burnell}},\ and\ \citenamefont
  {{Sondhi}}}]{sondhi}%
  \BibitemOpen
  \bibfield  {author} {\bibinfo {author} {\bibfnamefont {Y.}~\bibnamefont
  {{You}}}, \bibinfo {author} {\bibfnamefont {T.}~\bibnamefont {{Devakul}}},
  \bibinfo {author} {\bibfnamefont {F.~J.}\ \bibnamefont {{Burnell}}}, \ and\
  \bibinfo {author} {\bibfnamefont {S.~L.}\ \bibnamefont {{Sondhi}}},\
  }\href@noop {} {\bibfield  {journal} {\bibinfo  {journal} {arXiv e-prints}\
  ,\ \bibinfo {eid} {arXiv:1805.09800}} (\bibinfo {year} {2018})},\ \Eprint
  {http://arxiv.org/abs/1805.09800} {arXiv:1805.09800 [cond-mat.str-el]}
  \BibitemShut {NoStop}%
\bibitem [{\citenamefont {Pretko}\ and\ \citenamefont
  {Radzihovsky}(2018{\natexlab{a}})}]{leomichael}%
  \BibitemOpen
  \bibfield  {author} {\bibinfo {author} {\bibfnamefont {M.}~\bibnamefont
  {Pretko}}\ and\ \bibinfo {author} {\bibfnamefont {L.}~\bibnamefont
  {Radzihovsky}},\ }\href {\doibase 10.1103/PhysRevLett.120.195301} {\bibfield
  {journal} {\bibinfo  {journal} {Phys. Rev. Lett.}\ }\textbf {\bibinfo
  {volume} {120}},\ \bibinfo {pages} {195301} (\bibinfo {year}
  {2018}{\natexlab{a}})}\BibitemShut {NoStop}%
\bibitem [{\citenamefont {Pai}\ and\ \citenamefont {Pretko}(2018)}]{pai}%
  \BibitemOpen
  \bibfield  {author} {\bibinfo {author} {\bibfnamefont {S.}~\bibnamefont
  {Pai}}\ and\ \bibinfo {author} {\bibfnamefont {M.}~\bibnamefont {Pretko}},\
  }\href {\doibase 10.1103/PhysRevB.97.235102} {\bibfield  {journal} {\bibinfo
  {journal} {Phys. Rev. B}\ }\textbf {\bibinfo {volume} {97}},\ \bibinfo
  {pages} {235102} (\bibinfo {year} {2018})}\BibitemShut {NoStop}%
\bibitem [{\citenamefont {{Gromov}}(2017)}]{Gromov}%
  \BibitemOpen
  \bibfield  {author} {\bibinfo {author} {\bibfnamefont {A.}~\bibnamefont
  {{Gromov}}},\ }\href@noop {} {\bibfield  {journal} {\bibinfo  {journal}
  {ArXiv e-prints}\ } (\bibinfo {year} {2017})},\ \Eprint
  {http://arxiv.org/abs/1712.06600} {arXiv:1712.06600 [cond-mat.str-el]}
  \BibitemShut {NoStop}%
\bibitem [{\citenamefont {Pretko}\ and\ \citenamefont
  {Radzihovsky}(2018{\natexlab{b}})}]{pretko2018symmetry}%
  \BibitemOpen
  \bibfield  {author} {\bibinfo {author} {\bibfnamefont {M.}~\bibnamefont
  {Pretko}}\ and\ \bibinfo {author} {\bibfnamefont {L.}~\bibnamefont
  {Radzihovsky}},\ }\href@noop {} {\bibfield  {journal} {\bibinfo  {journal}
  {Physical review letters}\ }\textbf {\bibinfo {volume} {121}},\ \bibinfo
  {pages} {235301} (\bibinfo {year} {2018}{\natexlab{b}})}\BibitemShut
  {NoStop}%
\bibitem [{\citenamefont {Kumar}\ and\ \citenamefont
  {Potter}(2018)}]{kumar2018symmetry}%
  \BibitemOpen
  \bibfield  {author} {\bibinfo {author} {\bibfnamefont {A.}~\bibnamefont
  {Kumar}}\ and\ \bibinfo {author} {\bibfnamefont {A.~C.}\ \bibnamefont
  {Potter}},\ }\href@noop {} {\bibfield  {journal} {\bibinfo  {journal} {arXiv
  preprint arXiv:1808.05621}\ } (\bibinfo {year} {2018})}\BibitemShut {NoStop}%
\bibitem [{\citenamefont {Pretko}(2017{\natexlab{a}})}]{sub}%
  \BibitemOpen
  \bibfield  {author} {\bibinfo {author} {\bibfnamefont {M.}~\bibnamefont
  {Pretko}},\ }\href {\doibase 10.1103/PhysRevB.95.115139} {\bibfield
  {journal} {\bibinfo  {journal} {Phys. Rev. B}\ }\textbf {\bibinfo {volume}
  {95}},\ \bibinfo {pages} {115139} (\bibinfo {year}
  {2017}{\natexlab{a}})}\BibitemShut {NoStop}%
\bibitem [{\citenamefont {Pretko}(2017{\natexlab{b}})}]{genem}%
  \BibitemOpen
  \bibfield  {author} {\bibinfo {author} {\bibfnamefont {M.}~\bibnamefont
  {Pretko}},\ }\href {\doibase 10.1103/PhysRevB.96.035119} {\bibfield
  {journal} {\bibinfo  {journal} {Phys. Rev. B}\ }\textbf {\bibinfo {volume}
  {96}},\ \bibinfo {pages} {035119} (\bibinfo {year}
  {2017}{\natexlab{b}})}\BibitemShut {NoStop}%
\bibitem [{\citenamefont {Siva}\ and\ \citenamefont
  {Yoshida}(2017)}]{SivaYoshida}%
  \BibitemOpen
  \bibfield  {author} {\bibinfo {author} {\bibfnamefont {K.}~\bibnamefont
  {Siva}}\ and\ \bibinfo {author} {\bibfnamefont {B.}~\bibnamefont {Yoshida}},\
  }\href {\doibase 10.1103/PhysRevA.95.032324} {\bibfield  {journal} {\bibinfo
  {journal} {Phys. Rev. A}\ }\textbf {\bibinfo {volume} {95}},\ \bibinfo
  {pages} {032324} (\bibinfo {year} {2017})}\BibitemShut {NoStop}%
\bibitem [{\citenamefont {Prem}\ \emph {et~al.}(2017)\citenamefont {Prem},
  \citenamefont {Haah},\ and\ \citenamefont {Nandkishore}}]{prem}%
  \BibitemOpen
  \bibfield  {author} {\bibinfo {author} {\bibfnamefont {A.}~\bibnamefont
  {Prem}}, \bibinfo {author} {\bibfnamefont {J.}~\bibnamefont {Haah}}, \ and\
  \bibinfo {author} {\bibfnamefont {R.}~\bibnamefont {Nandkishore}},\ }\href
  {\doibase 10.1103/PhysRevB.95.155133} {\bibfield  {journal} {\bibinfo
  {journal} {Phys. Rev. B}\ }\textbf {\bibinfo {volume} {95}},\ \bibinfo
  {pages} {155133} (\bibinfo {year} {2017})}\BibitemShut {NoStop}%
\bibitem [{\citenamefont {Pai}\ \emph {et~al.}(2019)\citenamefont {Pai},
  \citenamefont {Pretko},\ and\ \citenamefont
  {Nandkishore}}]{pai2018localization}%
  \BibitemOpen
  \bibfield  {author} {\bibinfo {author} {\bibfnamefont {S.}~\bibnamefont
  {Pai}}, \bibinfo {author} {\bibfnamefont {M.}~\bibnamefont {Pretko}}, \ and\
  \bibinfo {author} {\bibfnamefont {R.~M.}\ \bibnamefont {Nandkishore}},\
  }\href {\doibase 10.1103/PhysRevX.9.021003} {\bibfield  {journal} {\bibinfo
  {journal} {Phys. Rev. X}\ }\textbf {\bibinfo {volume} {9}},\ \bibinfo {pages}
  {021003} (\bibinfo {year} {2019})}\BibitemShut {NoStop}%
\bibitem [{\citenamefont {Pretko}(2017{\natexlab{c}})}]{mach}%
  \BibitemOpen
  \bibfield  {author} {\bibinfo {author} {\bibfnamefont {M.}~\bibnamefont
  {Pretko}},\ }\href {\doibase 10.1103/PhysRevD.96.024051} {\bibfield
  {journal} {\bibinfo  {journal} {Phys. Rev. D}\ }\textbf {\bibinfo {volume}
  {96}},\ \bibinfo {pages} {024051} (\bibinfo {year}
  {2017}{\natexlab{c}})}\BibitemShut {NoStop}%
\bibitem [{\citenamefont {{Sala}}\ \emph {et~al.}(2019)\citenamefont {{Sala}},
  \citenamefont {{Rakovszky}}, \citenamefont {{Verresen}}, \citenamefont
  {{Knap}},\ and\ \citenamefont {{Pollmann}}}]{ah1}%
  \BibitemOpen
  \bibfield  {author} {\bibinfo {author} {\bibfnamefont {P.}~\bibnamefont
  {{Sala}}}, \bibinfo {author} {\bibfnamefont {T.}~\bibnamefont {{Rakovszky}}},
  \bibinfo {author} {\bibfnamefont {R.}~\bibnamefont {{Verresen}}}, \bibinfo
  {author} {\bibfnamefont {M.}~\bibnamefont {{Knap}}}, \ and\ \bibinfo {author}
  {\bibfnamefont {F.}~\bibnamefont {{Pollmann}}},\ }\href@noop {} {\bibfield
  {journal} {\bibinfo  {journal} {arXiv e-prints}\ ,\ \bibinfo {eid}
  {arXiv:1904.04266}} (\bibinfo {year} {2019})},\ \Eprint
  {http://arxiv.org/abs/1904.04266} {arXiv:1904.04266 [cond-mat.str-el]}
  \BibitemShut {NoStop}%
\bibitem [{\citenamefont {{Khemani}}\ and\ \citenamefont
  {{Nandkishore}}(2019)}]{ah2}%
  \BibitemOpen
  \bibfield  {author} {\bibinfo {author} {\bibfnamefont {V.}~\bibnamefont
  {{Khemani}}}\ and\ \bibinfo {author} {\bibfnamefont {R.}~\bibnamefont
  {{Nandkishore}}},\ }\href@noop {} {\bibfield  {journal} {\bibinfo  {journal}
  {arXiv e-prints}\ ,\ \bibinfo {eid} {arXiv:1904.04815}} (\bibinfo {year}
  {2019})},\ \Eprint {http://arxiv.org/abs/1904.04815} {arXiv:1904.04815
  [cond-mat.stat-mech]} \BibitemShut {NoStop}%
\bibitem [{\citenamefont {Ponte}\ \emph {et~al.}(2015)\citenamefont {Ponte},
  \citenamefont {Chandran}, \citenamefont {Papic},\ and\ \citenamefont
  {Abanin}}]{chandran}%
  \BibitemOpen
  \bibfield  {author} {\bibinfo {author} {\bibfnamefont {P.}~\bibnamefont
  {Ponte}}, \bibinfo {author} {\bibfnamefont {A.}~\bibnamefont {Chandran}},
  \bibinfo {author} {\bibfnamefont {Z.}~\bibnamefont {Papic}}, \ and\ \bibinfo
  {author} {\bibfnamefont {D.~A.}\ \bibnamefont {Abanin}},\ }\href {\doibase
  https://doi.org/10.1016/j.aop.2014.11.008} {\bibfield  {journal} {\bibinfo
  {journal} {Annals of Physics}\ }\textbf {\bibinfo {volume} {353}},\ \bibinfo
  {pages} {196 } (\bibinfo {year} {2015})}\BibitemShut {NoStop}%
\end{thebibliography}%

\end{document}